\documentclass{elsart}
\usepackage{graphicx}%  
\newcommand{\vol}[3]{\textbf{#1}\textrm{ (#3) #2}}
\newcommand{\EPJ} {Eur. Phys. J.}
\newcommand{\NIM} {Nucl. Inst. and Meth. in Phys. Res.}
\newcommand{\NP} {Nucl. Phys.}
\newcommand{\PL} {Phys. Lett.}
\newcommand{\PR} {Phys. Rev.}
\newcommand{\PREP} {Phys. Rep.}
\newcommand{\PRL} {Phys. Rev. Lett.}
\newcommand{\RPP} {Rep. Prog. Phys.}
\newcommand{\XeSn} {$^{129}$Xe+$^{\rm nat}$Sn}
\newcommand{\AuAu} {$^{197}$Au+$^{197}$Au}
\newcommand{\Etr} {$E_{\perp 12}$}

\begin{document}
\begin{frontmatter}

\title{Source shape determination with directional fragment-fragment 
velocity correlations}

\author[a]{A.~Le~F\`{e}vre},
\author[a]{C.~Schwarz},
\author[b]{G.~Auger},
\author[a]{M.L.~Begemann-Blaich},
\author[c]{N.~Bellaize},
\author[a]{R.~Bittiger},
\author[c]{F.~Bocage},
\author[d]{B.~Borderie},
\author[c]{R.~Bougault},
\author[b]{B.~Bouriquet},
\author[e]{J.L.~Charvet},
\author[b]{A.~Chbihi},
\author[e]{R.~Dayras},
\author[c]{D.~Durand},
\author[b]{J.D.~Frankland},
\author[d,k]{E.~Galichet},
\author[a]{D.~Gourio},
\author[f]{D.~Guinet},
\author[b]{S.~Hudan},
\author[f]{P.~Lautesse},
\author[d]{F.~Lavaud},
\author[e]{R.~Legrain\thanksref{dec}},
%\author[e]{R.~Legrain},
\author[c]{O.~Lopez},
\author[a,j]{J.~{\L}ukasik},
\author[a]{U.~Lynen},
\author[a]{W.F.J.~M\"{u}ller},
\author[e]{L.~Nalpas},
\author[a]{H.~Orth},
\author[d]{E.~Plagnol},
\author[g]{E.~Rosato},
\author[h]{A.~Saija},
\author[a]{C.~Sfienti},
\author[c]{B.~Tamain},
\author[a]{W.~Trautmann},
\author[i]{A.~Trzci\'{n}ski},
\author[a]{K.~Turz\'{o}},
\author[c]{E.~Vient},
\author[g]{M.~Vigilante},
\author[e]{C.~Volant},
\author[i]{B.~Zwiegli\'{n}ski}

\thanks[dec]{deceased}
\thanks{corresponding author: w.trautmann@gsi.de}

The INDRA and ALADIN Collaborations

\address[a]{Gesellschaft f\"{u}r Schwerionenforschung mbH, D-64291 Darmstadt,
Germany}
\address[b]{GANIL, CEA et IN2P3-CNRS, F-14076 Caen, France}
\address[c]{LPC Caen, ENSICAEN, Universit\'{e} de Caen, CNRS/IN2P3, F-14050 Caen, France}
\address[d]{Institut de Physique Nucl\'{e}aire, IN2P3-CNRS et Universit\'{e}, F-91406
Orsay, France}
\address[e]{DAPNIA/SPhN, CEA/Saclay, F-91191 Gif sur Yvette, France}
\address[f]{Institut de Physique Nucl\'{e}aire, IN2P3-CNRS et Universit\'{e}, F-69622
Villeurbanne, France}
\address[g]{Dipartimento di Scienze Fisiche e Sezione INFN, Univ. Federico II,
I-80126 Napoli, Italy}
\address[h]{Dipartimento di Fisica dell' Universit\`{a} and INFN, I-95129
Catania, Italy}

\address[i]{A.~So{\l}{}tan Institute for Nuclear Studies, Pl-00681 Warsaw, Poland}

\address[j]{H. Niewodnicza\'{n}ski Institute of Nuclear Physics, Pl-31342 Krak\'{o}w,
Poland}
\address[k]{Conservatoire National des Arts et M\'{e}tiers, F75141 Paris cedex 03, 
France}

\begin{abstract} 

Correlation functions, constructed from directional projections of the 
relative velocities of fragments, are used to determine the shape of the 
breakup volume in coordinate space. For central collisions of $^{129}$Xe + 
$^{\rm nat}$Sn at 50 MeV per nucleon incident energy, measured with the 4$\pi$ 
multi-detector INDRA at GSI, a prolate shape aligned along the beam direction
with an axis ratio of 1:0.7 is deduced. The sensitivity of the method 
is discussed in comparison with conventional fragment-fragment velocity 
correlations.

\end{abstract}


\end{frontmatter}

A recent study of central \XeSn ~and \AuAu ~collisions at bombarding 
energies between 50 and 100 MeV per nucleon has shown that a good 
statistical description of the measured fragment yields and kinetic 
energies can be obtained provided that a prolate source deformation 
and a superimposed collective motion are included \cite{lef04}. 
The experimental data had been collected with the 4$\pi$ INDRA 
multidetector \cite{pouthas} at the GSI laboratory. 
The statistical model employed in this study was 
the Metropolis Multifragmentation Monte-Carlo
(MMMC) model \cite{gross} which had been extended to non-spherical (NS)
sources~\cite{le_fevre}, a version referred to in the following as 
MMMC-NS model. The MMMC Statistical Model is based on the microcanonical 
ensemble and has found many applications in nuclear multifragmentation
(see, e.g., \cite{gross,sneppen,baoan94,marieluise,botmish06}).

A prolate deformation of the emitting source along the direction of the 
incident beam was indicated by the observed anisotropies of the fragment 
production. The element spectra were found to extend to larger atomic numbers 
$Z$ at forward and backward emission angles than at sideward angles. 
The largest fragment within an event is preferentially emitted toward 
forward or backward angles.
If emitted sidewards, its mean $Z$ decreases, e.g., from
$Z \approx$ 18 at $\theta_{\rm cm}$~=~0$^{\circ}$ to $Z \approx$ 14 at 
$\theta_{\rm cm}$ = 90$^{\circ}$ 
for the 1\% most 
central collisions of \XeSn ~at 50 MeV per nucleon, selected on the basis
of the measured charged-particle multiplicity or the transverse energy of
light charged particles~\cite{lef04}.

In the model description, an important role is played by the Coulomb 
interaction which favours large separations between heavy fragments in 
order to minimize the Coulomb energy. Heavy fragments are, therefore,
preferentially placed in the tips of a prolate source. Coulomb repulsion and 
the superimposed radial flow
transform these spatial correlations into correlations in momentum space
which produces the observed maxima in the yields and kinetic energies of the
heaviest fragments at forward and backward directions. 
The orientation along the beam axis is clearly of a dynamical origin. The
question, therefore, arises whether the deformation of the model source,
besides the superimposed flow, 
is mainly required for simulating the observed 
anisotropies in momentum space or whether 
it actually reflects the source shape at breakup as caused by the reaction 
dynamics.

It is the aim of the present work to confirm the indicated source 
deformations in coordinate space 
by applying interferometric methods to the same data,
here at first only for the case of \XeSn ~ at 50 MeV per nucleon.
Interferometry has become a standard tool for investigating the 
space-time properties of the breakup state in heavy-ion reactions 
\cite{ardouin,wiedemann,verde06}. 
The spatial dimensions or space and time are usually separated
by exploiting the effects of quantum statistics in proton-proton, 
neutron-neutron, or pion-pion correlation functions 
\cite{wiedemann,koonin,lisa93,colonna95,heinz99,schwarz01}. 
Particle pairs are selected according to the orientation of their
relative velocity with respect to the chosen coordinate axes,
and separate correlation functions are generated as, e.g., for longitudinal
and transverse orientations.
In fragmentation reactions, the mutual Coulomb repulsion between fragment 
pairs has been used to derive time scales for the emission 
process from fragment-fragment correlation functions 
\cite{trockel,fox94,beaulieu00,rodionov02,taba06}.
Glasmacher et al. have shown that also information on the source 
geometry can be obtained if directional cuts are applied 
and correlation functions are generated for fragment pairs 
with relative velocities parallel or perpendicular 
to the symmetry axis \cite{glas93,glas94}. 
The dependence on the source geometry was 
small, even though nuclei with the unusual shapes 
of flat disks or toroids were included in these studies. 
It will be shown in the following that the sensitivity 
to the source geometry is significantly enhanced with the  
proposed new kind of correlation functions of projections.

The new correlation functions are constructed from the relative velocities
of fragment pairs by not using their modulus, as for conventional fragment-fragment
correlations, but rather their longitudinal and transverse components with
respect to the direction of orientation which here is the beam axis. 
To correct, in first order, for the variation of the Coulomb repulsion with
the fragment $Z$, the reduced velocity 
$v_{\rm red}=v_{\rm rel}/\sqrt{Z_{i}+Z_{j}}$ is 
used, where $v_{\rm rel}$, $Z_{i}$ and $Z_{j}$ are the relative velocity and the 
atomic numbers of the two fragments, respectively \cite{kim92}. 
With the help of model calculations, it will be shown that these 
projected correlations are particularly sensitive to a deformation 
of the source. This will be done in comparison to conventional correlation 
functions generated not with directional cuts but with
directional weights, proportional to sin$^2\theta$ and cos$^2\theta$
where $\theta$ is the polar angle of the relative velocity vector 
with respect to the beam axis \cite{schwarz01}.
The method will then be applied to the
experimental data for the \XeSn ~reaction at 50 MeV per nucleon.

\begin{figure}[!htb]	%Fig. 1
%\vspace{-3mm}
    \leavevmode
\begin{center}

\includegraphics[height=11.0cm]{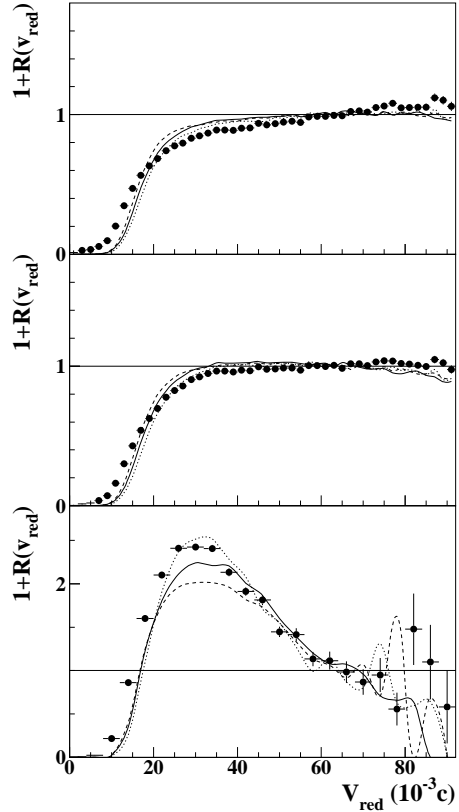}
\end{center}
%     \epsfysize=11.0cm
%   \epsffile{Fig1.eps}

    \caption{Fragment-fragment velocity correlations as a function of
the reduced velocity $v_{\rm red}$ (in units of $10^{-3} c$) of fragment 
pairs with $3 \leq Z \leq 30$ (top and middle panels) and of the two 
fragments with the 
largest $Z$ within an event (bottom panel) for central collisions
of \XeSn ~at 50 MeV per nucleon (dots). 
Solid, dashed and dotted lines represent the filtered predictions
of the MMMC-NS model 
for the prolate, oblate and spherical sources, respectively.
%The correlation functions shown in the middle and bottom panels are generated in the coordinate frame of the kinetic-energy tensor spanned by all detected charged particles and fragments. 
The uncorrelated denominator was generated after azimuthal event alignment for 
the top panel and after polar event alignment for the middle and 
bottom panels (see text).
The interval $50-100 \cdot 10^{-3} c$
is chosen for normalization.
}

\label{fig:velcorr} 
\end{figure}  

Experimental details of these measurements, performed at the GSI laboratory 
in 1998 and 1999, and of the analysis and calibration procedures 
may be found in \cite{lef04,lukasik02,trzcinski,turzo04}. %turzo ?
Natural Sn targets of areal density 1.05 mg/cm$^2$ were bombarded with 
$^{129}$Xe beams delivered by the SIS heavy-ion synchrotron. 
Central impact parameters were selected by requiring that the total 
transverse energy \Etr ~of light charged particles ($Z$ = 1,2) 
exceeded 637 MeV \cite{lef04}. 
This corresponds to an impact-parameter range $b/b_{\rm max} \leq 0.1$, 
according to the geometrical prescription of \cite{cavata}. 

\begin{figure}[!htb]		%Fig. 2
%\vspace{-3mm}
    \leavevmode
\begin{center}

\includegraphics[height=10.0cm]{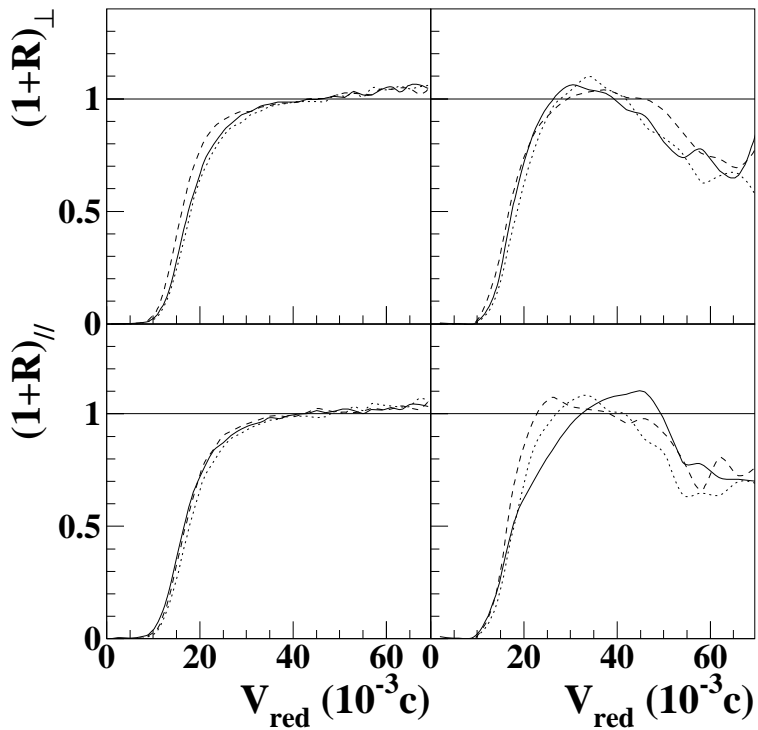}
\end{center}

%    \centering
%     \epsfysize=9.0cm
%   \epsffile{Fig2_bis.eps}

    \caption{Directional correlation functions obtained from filtered 
predictions of the MMMC-NS model for 
fragment pairs with $3 \leq Z \leq 30$ (left panels) and for the two 
fragments with largest $Z$ within each event (right panels). 
The top and bottom panels show the transverse and longitudinal 
correlation functions, respectively, obtained with the method of  
directional weights \protect\cite{schwarz01}. 
Solid, dashed and dotted lines represent the predictions 
for the prolate, oblate and spherical sources, respectively.
The interval $30-100 \cdot 10^{-3} c$ is chosen for normalization.
}

\label{fig:weicorr} 
\end{figure}

\begin{figure} [!htb]		%Fig. 3
    \leavevmode
\begin{center}

\includegraphics[height=11.0cm]{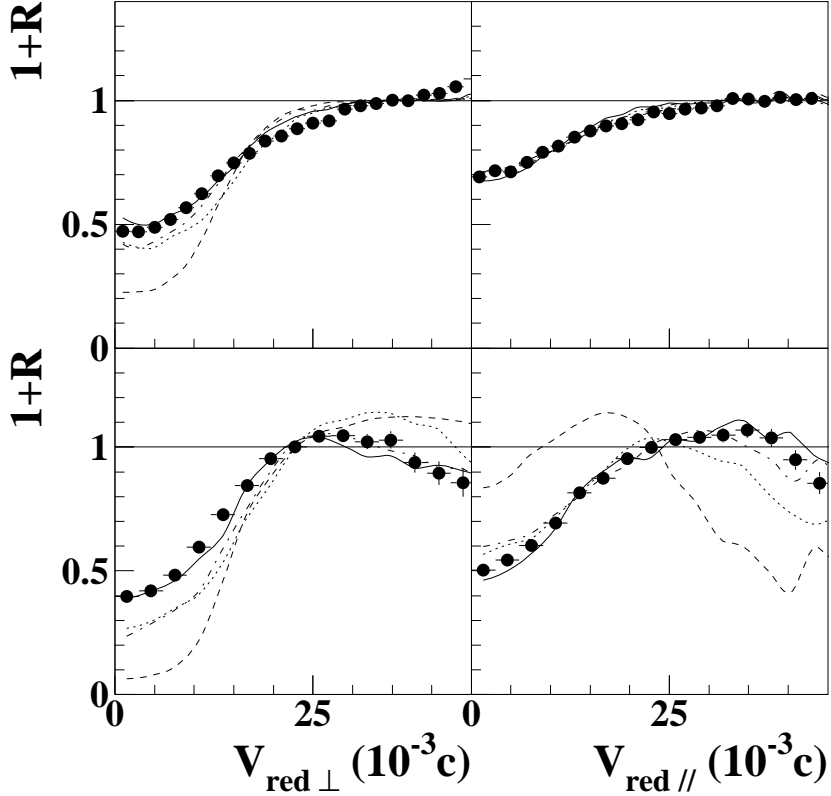}
\end{center}

%    \centering
%     \epsfysize=9.0cm
%   \epsffile{Fig3.eps}

    \caption{Correlation functions constructed from the transverse 
(left panels) and longitudinal (right panels) projections of the 
reduced relative velocity of fragment pairs with 
$3 \leq Z \leq 30$ (top panels) 
and of the two fragments with largest $Z$ (bottom panels)
for central \XeSn ~collisions at 50~MeV per nucleon (symbols). 
Solid, dashed and dotted lines represent the filtered 
predictions of the MMMC-NS model for the prolate, 
oblate and spherical sources, respectively. The dashed-dotted lines
(closely following the dotted lines at small $v_{\rm red}$) represent
the spherical source with ellipsoidal flow 
(see Ref.~\protect\cite{lef04}).
The normalization was performed, individually for each projection,
within $30-40 \cdot 10^{-3} c$ in the top 
and $20-30 \cdot 10^{-3} c$ in the bottom panels, respectively.
}

\label{fig:procorr}
\end{figure}

\begin{figure} [!htb]		%Fig. 4
    \leavevmode
\begin{center}

\includegraphics[height=6.5cm]{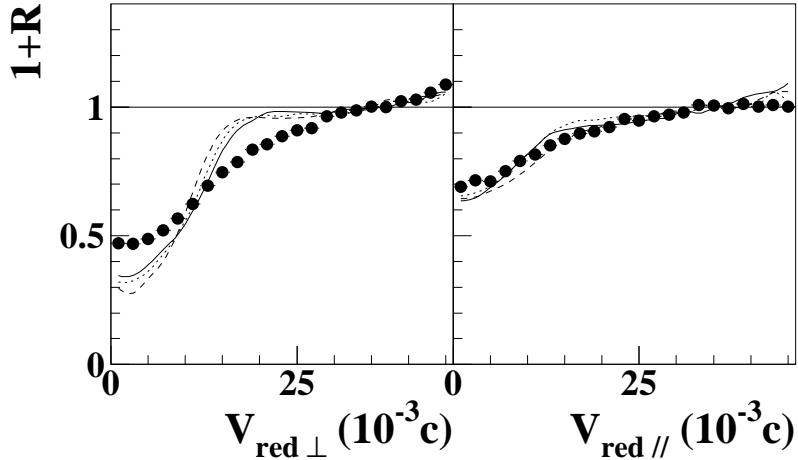}
\end{center}

%    \centering
%     \epsfysize=5.8cm
 %  \epsffile{Fig4.eps}

    \caption{Correlation functions constructed from the transverse 
(left panel) and longitudinal (right panel) projections of the 
reduced relative velocity of fragment pairs with $3 \leq Z \leq 30$
for central \XeSn ~collisions at 50~MeV per nucleon (symbols, same data
as in Fig.~\protect\ref{fig:procorr}, top panels). 
Solid, dashed and dotted lines represent the predictions 
of the MMMC-NS model without radial flow for the prolate, 
oblate and spherical sources, respectively.
The normalizations were performed within $30-40 \cdot 10^{-3} c$.
}

\label{fig:noflow}
\end{figure}

Standard fragment-fragment velocity correlations constructed 
for this event class are shown in Fig.~\ref{fig:velcorr}.
They were obtained with the technique of event mixing, and 
the interval $50-100\cdot 10^{-3} c$ was chosen for normalization
($c$ is the velocity of light). In the top panel, 
the uncorrelated denominator was 
generated after rotating all events, azimuthally with
respect to the beam axis, into a common reaction plane in order to suppress 
artificial flow effects \cite{kotte95}.
The orientation of an event was determined from that of the
longest eigenvector of the kinetic-energy tensor, built from all detected
charged particles. 
A flat correlation function at large $v_{\rm red}$ is obtained by aligning
all events with respect to each other before generating the uncorrelated 
denominator, e.g. by rotating their longest eigenvectors into the beam direction 
(Fig.~\ref{fig:velcorr}, middle panel). 
With this polar event alignment, the distribution of mixed pairs extends to 
slightly larger $v_{\rm red}$, leading to a reduction of $1+R$ with respect 
to the case of azimuthal alignment only (top panel). The simpler 
method of azimuthal event alignment has, nevertheless, been used in the 
analysis described below (Figs.~\ref{fig:weicorr}-\ref{fig:noflow}).

All correlation functions are characterized by a strong suppression 
of small relative velocities, with $1+R$ falling below 0.5 for 
$v_{\rm red} \leq 14\cdot 10^{-3} c$.
Similarly large suppressions have been observed 
for a variety of heavy-ion and light-ion induced reactions
and were consistently interpreted as indicating short breakup times of the
order of 50 - 100 fm/$c$~\cite{rodionov02,glas94,lopez,wang}.
The correlation function of the pairs with largest $Z$ in each event
(Fig.~\ref{fig:velcorr}, bottom panel) 
exhibits a considerable enhancement around $30\cdot 10^{-3} c$.
With mean values $\langle Z_1 \rangle=13$ for the largest and $\langle Z_2 \rangle=8$ 
for the second largest fragment~\cite{lef04}, 
the bump corresponds to relative velocities of $\approx 4$ cm/ns, 
a value typically observed for large-angle
fragment-fragment correlations \cite{baoan94,pocho89}. The distribution of
relative angles between the two largest fragments in the center-of-mass 
system is indeed wide and has a mean value of $\approx 100^{\circ}$.

The MMMC-NS model \cite{le_fevre} has been used as an event
generator for constructing correlation functions to be 
compared with the experimental data. 
For the global source parameters the previously obtained values 
of total charge $Z = 79$ and mass $A = 188$, 
excitation energy $E_x =$ 6 MeV per nucleon, 
and collective flow 2 MeV per nucleon with a profile parameter
$\alpha_{\rm coll} = 2$ were used \cite{lef04}. 
Only the shape parameters were varied, and three different source 
elongations were chosen, a prolate source with axis ratio 1:0.70, 
an oblate source with axis ratio 1:1.67, and a spherical source.
The produced MMMC-NS events have been filtered using a software replica
of the experimental apparatus that takes into account the main properties 
of the INDRA multidetector as, e.g., 
the exact geometry of the detectors, the energy thresholds and limits for 
charged-particle detection and identification, and the effects of the 
multihit treatment in the off-line analysis.

The standard correlation functions depend very weakly on the source shape 
(Fig.~\ref{fig:velcorr}). Only the correlation function for the two largest
fragments exhibits slightly different maxima for deformed sources whose
normalization, however, in the absence of an uncorrelated plateau region, 
is less certain (Fig.~\ref{fig:velcorr}, bottom panel).
The depression at small $v_{\rm red}$ is well reproduced, even though the
predicted rise around $v_{\rm red} = 15\cdot 10^{-3} c$ is steeper than 
found experimentally. Smoother correlation functions could probably be 
generated and a better agreement reached with finite emission times and by 
allowing the source parameters to fluctuate (cf.~Ref.~\cite{kim92a}). 
 
The dependence of conventional 
directional fragment-fragment correlation functions on
the shape of the emitting source in coordinate space is illustrated in 
Fig.~\ref{fig:weicorr}. Longitudinal and transverse correlation functions 
were generated with the method of directional weights \cite{schwarz01}, 
i.e. by accumulating the numerators and denominators
with weight functions cos$^2\theta$ and sin$^2\theta$, respectively. 
Their dependence on the source shape is visible but weak, 
similar to the results reported in
\cite{glas93,glas94}. 

A much stronger sensitivity to the deformation is exhibited by 
the correlation functions generated separately for the longitudinal 
and transverse projections of the relative velocity $v_{\rm red}$ 
(Fig. \ref{fig:procorr}). Note that for the projections shown here and below 
different intervals were chosen for normalization (see captions). 
Since individual components can become small even if the modulus of
$v_{\rm red}$ is large, 
these correlation functions do not approach zero at small values of the 
projected $v_{\rm red}$. Their magnitude there is sensitive to correlations of 
the orientations of the relative-velocity vectors which, apparently, depend
strongly on the source shape. The correlations of the transverse projections
at small $v_{\rm red \perp}$, in particular, change significantly with the source 
geometry. As will be shown and 
discussed below, the sensitivity to radial flow is large as well. 
With the flow value 2 MeV per nucleon determined from the fragment 
kinetic energies, a good agreement with the experimental data for 
\XeSn ~is reached by assuming a prolate deformation.
This is valid for the correlation functions generated from all pairs with
$Z_{i,j} > 2$ but, in particular, also for the correlation functions for 
the two largest 
fragments of each partition (Fig. \ref{fig:procorr}, bottom). 
The comparison thus confirms the conclusions drawn from the study of 
the fragment anisotropies 
in yields and kinetic energies \cite{lef04}. 

Projected correlation functions generated with the MMMC model for central
\XeSn ~collisions with the same input parameters but with the 
collective radial flow set to zero are shown in Fig.~\ref{fig:noflow}.
Without flow, their rise at small $v_{\rm red}$ is much steeper, 
notably for the transverse projections, and their differences 
for the three considered source shapes are much smaller.
It is evident that the experimental data for \XeSn ~are qualitatively 
different from this  
set of correlation functions and cannot be reproduced without assuming 
collective contributions to the particle and fragment energies.

\begin{figure} [!htb]		%Fig. 5
    \leavevmode
\begin{center}

\includegraphics[width=12.5cm]{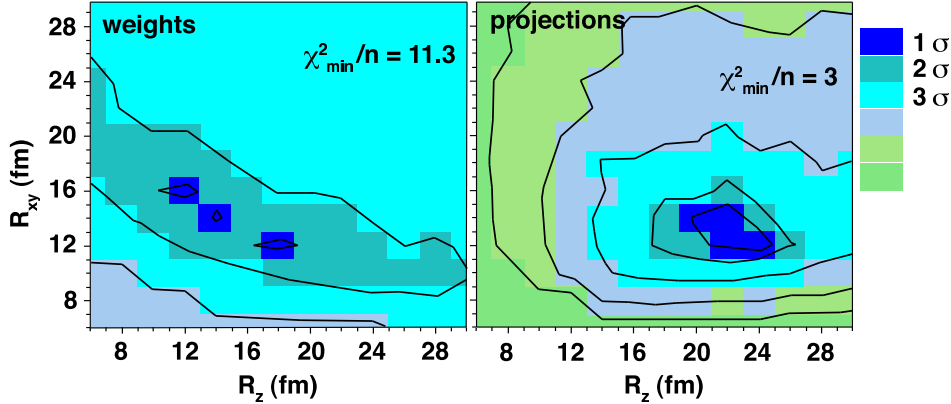}
\end{center}

%    \centering
%     \epsfysize=10.0cm

%   \epsffile{carsten1-2.eps}
%   \epsffile{Fig5.eps}
\vspace{3mm}

  \caption{(color online) 
Contour plots (in units of the standard deviation $\sigma$) of the
two-dimensional $\chi^2$ distributions in the plane of the transverse (xy)
and longitudinal (z) extensions of the source, as obtained from the 
comparison of the experimental weighted (left panel) and projected (right panel) 
correlation functions, for fragments with $Z=5-7$, with the predictions 
of the classical deformed-source model for \XeSn ~at 50 MeV per nucleon.
}
\vspace{-1mm}

\label{fig:chi2_CS}
\end{figure}

The deduced sensitivities to the source shape in coordinate space were
independently verified by reproducing the experimental 
correlation functions with a simpler source model \cite{lef_bormio03}.
Fragments were sampled according to the experimental yield distribution 
and placed randomly without overlap into a predetermined ellipsoidal volume
with the transverse ($R_x, R_y$) and longitudinal ($R_z$) source radii being 
varied between 6 fm and 30 fm. 
The parameters describing the thermal and anisotropic collective motions 
of the fragments were chosen, 
individually as a function of $Z$,
% and varying with the source deformation, nein!!! (Carsten)
so as to best reproduce the measured kinetic energies. The temporal evolution 
after freeze-out was modeled with N-body Coulomb trajectory calculations. 
The asymptotic velocities were used to construct both, weighted and projected, 
correlation functions. The result of a comparison with the experimental data, 
for intermediate-mass fragments with $5 \leq Z \leq 7$, is shown in 
Fig. \ref{fig:chi2_CS} in the form of the $\chi^2$ distributions in the plane of the 
transverse and longitudinal extensions, representing the degree of agreement obtained.

The different sensitivities of the two types of correlation functions are
immediately apparent. With directional weights, the minima
of the $\chi^2$ distribution extend over a broad band of radius values, 
all corresponding to approximately the same volume but to different shapes. 
For the correlation functions constructed from the projections 
(right panel of Fig.~\ref{fig:chi2_CS}), the minima are concentrated in a 
region of prolate source shapes with axis ratios of $1:0.6\pm0.1$.
Other observations were also found to be similar to those made with the more complex 
MMMC model. In particular, the requirement of a finite radial flow for
obtaining the shape dependence of the projected correlation functions was
confirmed. Radial flow leads to a correlation between the locations of a particle
in coordinate space and in momentum space which makes its role here 
understandable.
However, flow by itself is not sufficient (cf. dotted and dashed-dotted
lines in Fig.~\ref{fig:procorr}). The projected correlation functions become
significantly different only in the presence of deformations 
in coordinate space (full and dashed-dotted lines in Fig.~\ref{fig:procorr}).

The volumes obtained with this method are large. % but not unreasonable. 
A mean radius of 15 fm (Fig.~\ref{fig:chi2_CS}, left panel) corresponds to 
about twice the radius of a system of $A=$~188 nucleons at normal density, 
slightly larger than
what is usually assumed in applications of the MMMC model 
(6 times the normal nuclear volume in Refs.~\cite{lef04}) or what is obtained 
with other methods~\cite{parlog05,piant05}. A very similar volume results from
the transverse and longitudinal radii of 13 fm and 22 fm, respectively, favoured
by the method of projections (Fig.~\ref{fig:chi2_CS}, right panel).
Considering the finite range of $Z$ used for this comparison, this is
perhaps not unreasonable. %this seems rather satisfactory. 
It is, furthermore, known that large variations
of the volume are necessary in order to change fragment-fragment correlation 
functions significantly \cite{schapiro}.
An additional contribution to the apparent source size may also come from the finite
time scale of the emission process even though it is expected to be rather 
short, perhaps 50 - 100 fm/$c$ (see, e.g.,~Ref.~\cite{glas94,zbiri07}). 
Estimates for the increase of the mean source radius, 
obtained from simulations with the above-mentioned simpler 
source model using finite
emission times, are of the order of 1 - 2 fm. 
They are non-negligible regarding the corresponding breakup volume 
but they are unlikely to change the parameters describing its anisotropy. 

In summary, a new method based on correlation functions of relative-velocity
projections has been presented and shown to be sensitive to source deformations 
in the presence of flow.
It has been applied to the data for central collisions of \XeSn ~at 50 MeV 
per nucleon, and an expanded source with a prolate elongation in the direction of the 
incident beam has been deduced. The axis ratio $1:0.7$ used for the quantitative 
comparison with the MMMC-NS model predictions has led to very satisfactory results,
thus supporting the assumption of a non-spherical breakup source with a deformation 
%thus validating the assumption of a non-spherical breakup source with a deformation 
of this magnitude 
%in good agreement with the deformation of the  
%non-spherical breakup source required  
for the statistical description of the fragmentation channels with this 
model~\cite{lef04}. A similar deformation of $1:0.6\pm0.1$ has resulted from the
analysis with a simpler source model, restricted to the $Z$ = 5-7 range of fragments. 
	  
The dynamical origin of this deformation was not a subject of the present study
but is most likely found in the
incomplete mutual stopping of the incident ions, even in the most violent 
collisions associated with the largest transverse energies or particle 
multiplicities. Stopping is incomplete even in the heavier 
\AuAu ~system for which an excitation function of stopping 
with a maximum at several hundred MeV per nucleon has been established 
\cite{reisdorf04,andronic06}. At the present energy of 50 MeV per nucleon 
and below, the observed mass hierarchy of the longitudinal fragment
velocities provides further evidence for dynamical correlations between 
the entrance and exit channels~\cite{colin03}. 
In coordinate space, the larger longitudinal momenta will cause 
the source shape to be elongated at the time when the breakup density has 
been reached. Larger fragments, presumably carrying most of the 
remaining longitudinal momentum, will be near the tips of 
the elongated source, an effect which in the static  
MMMC-NS description is generated by the Coulomb force alone.
Coulomb effects are probably less dominant in the dynamical situation which 
explains why the deformed-source model, by its construction, permits  
exploring the statistical nature of multi-fragment breakups 
in the presence of dynamical constraints which are not part 
of the model. Source deformations and anisotropic emissions are 
generic features of central collisions at intermediate energies, a fact already
quite well established but supported in a new way with the present method. 
Since the radial flow is increasing it should be possible to extend these 
studies to higher bombarding energies.

This work was supported 
by the European Community under
contract No. ERBFMGECT950083.

\end{document}